\documentstyle[preprint,eqsecnum,cite,epsf,aps,amssymb]{revtex}

\newcommand{\be}{\begin{equation}}
\newcommand{\ee}{\end{equation}}
\newcommand{\bea}{\begin{eqnarray}}
\newcommand{\eea}{\end{eqnarray}}
\newcommand{\ba}{\begin{array}}
\newcommand{\ea}{\end{array}}
\newcommand{\nn}{\nonumber \\}
\newcommand{\half}{\frac{1}{2}}

\newcommand{\Z}{\mathbb{Z}}
\newcommand{\R}{\mathbb{R}}

\begin{document}

\title{Off-Shell Duality in Born-Infeld Theory}

\author{Cedric R. Le\~ao and Victor O. Rivelles}

\address{Instituto de F\'\i sica, Universidade de S\~ao Paulo\\
 Caixa Postal 66318, 05315-970, S\~ao Paulo - SP, Brazil\\
E-mail: cedric@fma.if.usp.br, rivelles@fma.if.usp.br}

\maketitle

\begin{abstract}
The classical equations of motion of Maxwell and Born-Infeld theories
are known to be invariant under a duality symmetry acting on the field
strengths. We implement the $SL(2,\Z)$ duality in these
theories as linear but non-local transformations on the potentials. We
show that the action and the partition function in the Hamiltonian
formalism are modular invariant in any gauge. For
the Born-Infeld theory we find that the longitudinal part of the
fields have to be complexified. 
\end{abstract}

\newpage


\section{INTRODUCTION}
\label{I}

Duality plays a fundamental role in describing the same 
physical system using different variables. It provides a valuable tool
to understand different aspects of the same theory. For instance, the five
perturbative string theories are now known to be related
to each other through a series of different dualities. Also, D-branes
are known to be solitonic objects in string theory and as such they
carry information about its non-perturbative sector. When 
string or M-theory is consistently truncated the resulting quantum
field theory also presents some duality which is reminiscent from that
of the larger theory. 
Some supersymmetric gauge theories inherit such dualities as, for
instance, the ${\cal N}=4$ supersymmetric Yang-Mills theory. A similar
situation occurs for D-branes whose low energy limit is described by a
Born-Infeld theory. In particular, the D3-brane action of type IIB
string theory is $SL(2,\Z)$ self-dual and this symmetry is 
inherited by the four dimensional Born-Infeld theory
\cite{Tseytlin}. Usually the original theory and the dual one are
distinct, but sometimes the dual theory coincides with the original
one. We will concentrate on such theories in this paper. 

In field theory duality is often realized as symmetries of the
equations of motion. It is well known that the equations of motion,
Bianchi identities and the energy-momentum tensor of Maxwell and
Born-Infeld theories are invariant under $SO(2)$  
rotations which mix the electric and magnetic fields
\cite{Gibbons-Rasheed,Gaillard-Zumino}. The action, 
however, is not invariant. The $SO(2)$ symmetry
can be enlarged to $SL(2,\R)$ when a dilaton and an axion are added
\cite{Gibbons-Rasheed2, Gaillard-Zumino}. This motivated the search
for gauge theories whose equations of motion are duality invariant. 

As stated before, duality is found as symmetries of the classical 
equations of motion and energy-momentum tensor. However, it is
desirable that such symmetries could be implemented at the quantum
level as symmetries of the action and partition function.
In this way the symmetries will hold in any situation and not only for
on-shell quantities. This is relevant, for instance, in the derivation
of Ward identities among the off-shell Green's functions. Off-shell
symmetries must be implemented in the 
basic field variables (and not in the field strengths for gauge
theories) either in the Lagrangian or Hamiltonian
formalism. When varying the action the resulting boundary term must be
local in time, giving rise to a Noether current associated to the
invariance. However, the boundary term can be non-local in space
provided that it has a sufficient falloff at spatial infinity. 
This allows  the variations of the basic field variables to be
non-local in space. These ideas were
first explored in \cite{Deser-Teitelboim} where the $SO(2)$ symmetry of
Maxwell equations were implemented at the 
action level in the Hamiltonian formalism in Coulomb gauge. The
transformations of the vector potential 
and its canonical momentum are non-local in space
\be
\label{1}
\delta A_i = \alpha  \epsilon_{ijk} \frac{\partial^j}{\nabla^2} E^k,
\qquad \delta E_i = \alpha \epsilon_{ijk} \partial^j A^k.
\ee
On-shell they give rise to the usual $SO(2)$ transformation between
the electric and magnetic fields $\delta E_i = \alpha B_i, 
\delta B_i = - \alpha E_i$. The corresponding Noether charge, 
which generates the rotation, is also non-local and has an expression
involving Chern-Simons terms \cite{Deser-Teitelboim}. The same holds
for the Born-Infeld theory \cite{Bengtsson} and for gauge theories
coupled to matter and gravity \cite{Igarashi}.

The transformations Eqs.(\ref{1}) leave the action invariant only in
the Coulomb gauge. This could be seen as a drawback since the symmetry
manifests itself only in a particular gauge. Even so, it may be quite
useful. A typical example is the Chern-Simons theory in Landau
gauge. In this case there appears a vector supersymmetry
\cite{Chern-Simons} which can be extended to the exceptional algebra
$D(2,\alpha)$ \cite{Damgaard}. This symmetry is essential to show the
renormalizability of the model. 

We should point out that there is an alternative procedure to
implement off-shell symmetries in the action with local transformation
laws. Usually it breaks manifest Lorentz invariance and demands the
introduction 
of more fields. For the Maxwell theory this requires a description in
terms of two potentials giving rise to the Schwarz-Sen model
\cite{Schwarz-Sen} or, alternatively, an infinite number of them 
\cite{McClain}. Duality manifests itself as rotations between the
potentials. It is possible to show that the duality symmetry of the
Schwarz-Sen model is the local form of the non-local transformations  
Eqs.(\ref{1}) \cite{Girotti-Gomes-Rivelles}. Although the Schwarz-Sen
model is not manifestly Lorentz covariant this symmetry can be made
manifest by the inclusion of auxiliary fields and some gauge symmetry
through 
the PST formalism \cite{PST}. A similar situation is found for the
Born-Infeld theory \cite{Parra}. It should be remarked that
this situation is 
not exclusive of duality symmetry. Even well known symmetries, like the
BRST symmetry, can be cast into a non-local form at the expense of
loosing manifest Lorentz invariance \cite{Rivelles}. In this work,
however, we shall not follow this approach. 

The $SL(2,\R)$ symmetry of the equations of motion found when a
dilaton and an axion are added, manifests, at the quantum level, as an
$SL(2,\Z)$ duality of the partition function. This happens when the
dilaton and the axion take their vacuum expectation value which are
combined into a complex coupling constant $\tau$ with its real part
being the theta term. Now the action and the partition function are
functions of $\tau$ and duality manifests as modular transformations
of the coupling constant\footnote{A function $F(\tau)$ is said to be a
  modular form of weight $(u,v)$ if, under a modular transformation
  $\tau \rightarrow \tilde{\tau}$, Eq.(\ref{3.2}) below, it transforms
  as $F(\tilde{\tau}) = (c \tau + d)^u (c \tilde{\tau} + d)^v
  F(\tau)$.}. There are 
two basic ways to implement duality in gauge theories. In the first one a
new gauge field is introduced in such a way that upon functional
integration over the original field a dual theory is obtained
\cite{Witten}. The second way treats duality as a canonical
transformation of the original phase space variables \cite{Lozano}. In
this paper we will concentrate in 
this second approach since it can be connected with the classical
symmetries of the equations of motion. 

In Maxwell theory the Lagrangian partition function is found to be a
modular form under $SL(2,\Z)$ transformations of the coupling
constant $\tau$ \cite{Witten,Olive-Alvarez}. The weights of the
modular transformation are proportional 
to the Euler number and the signature of the space-time. At the
Hamiltonian level, the partition function is modular invariant with
modular weight equal to zero \cite{Lozano,Olive-Alvarez}. In this case
duality can be implemented as a canonical transformation on the
reduced phase space. This means that Gauss law holds and we are
on-shell. Also, the canonical transformation has 
essentially the form Eq.(\ref{1}) and holds only in Coulomb gauge. 

In this paper we will show how the non-local $SO(2)$ transformations
Eqs.(\ref{1}) can be extended to $SL(2,\Z)$ transformations for the
Maxwell and Born-Infeld theories with a theta term. 
We will also show that duality holds off-shell in the sense that Gauss
law is not required. It holds also in any gauge,
and not just in Coulomb gauge as originally proposed in
\cite{Deser-Teitelboim}. We will find that the generalization of
Eqs.(\ref{1}) to the $SL(2,\Z)$ case is also non-local in space. To
study duality at the quantum level we consider the phase
space partition function and the Hamiltonian BRST formalism. Since the
Born-Infeld theory is non-renormalizable we treat it as an effective
field theory and its partition function should be considered in this
context. We will show that the $SL(2,\Z)$ transformations can be
regarded as a canonical transformation and that the phase space
partition function is modular invariant. 
We also find that in order to implement a linear $SL(2,\Z)$
transformation for the Born-Infeld theory it is necessary to consider
the longitudinal part of the fields as being complex.

\section{$SO(2)$ Duality in Maxwell Theory}
\label{II}

Maxwell theory with a theta term is described by the following action
in Minkowski space with metric $(+---)$ 
\be
\label{2.1}
S = - \frac{1}{8 \pi} \int d^4x \,\, \left( \frac{4 \pi}{g^2} F^{\mu\nu}
F_{\mu\nu} + \frac{\theta}{2\pi} F^{\mu\nu} {}^{*}F_{\mu\nu} \right),
\ee
where ${}^{*}F^{\mu\nu} = \frac{1}{2} \epsilon^{\mu\nu\rho\sigma}
F_{\rho\sigma}$. The Hamiltonian formulation is obtained in a
straightforward way. There is a primary constraint $\Pi^0 =
\frac{\delta L}{\delta \dot{A}_0} = 0$ and the secondary constraint
receives no contribution from the theta term, giving rise to the usual
Gauss law $\partial_i \Pi^i=0$. The Hamiltonian density is then  
\be
\label{2.2}
H_M = - \frac{2 \pi i}{\tau - \overline{\tau}} \Pi^i \Pi_i - i
\frac{\tau + \overline{\tau}}{\tau - \overline{\tau}} \Pi^i B_i -
\frac{i}{2\pi} \frac{\tau \overline{\tau}}{\tau - \overline{\tau}} B^i
B_i,
\ee
where $\Pi^i = \frac{\delta L}{\delta \dot{A}_i}$, the magnetic field
is $B_i = \frac{1}{2} \epsilon_{ijk}F^{jk}$ 
and the complex coupling constant is $\tau=\frac{\theta}{2\pi} +
\frac{4\pi i}{g^2}$. The contribution from the theta term appears in the
second and third terms of Eq.(\ref{2.2}). 

The BRST charge is the same as in pure Maxwell theory since the
constraint structure was not modified 
\be
\label{3}
Q = \int d^3x \,\, \left( \partial_i \Pi^i C + {\cal P}_D \Pi_0 \right).
\ee
The ghosts obey the canonical Poisson brackets $\{ {\cal P}_C, C
\} = \{ {\cal P}_D, D \} = -1$ and the BRST transformations
are
\bea
\label{4}
&& \delta A_i = \partial_i C,  \qquad \delta A_0 = - {\cal P}_D, \qquad
\delta \Pi_0 = 0,  \qquad \delta \Pi_i = 0, \nn
&& \delta C= 0, \qquad\quad\,  \delta {\cal P}_C = - \partial_i \Pi^i,
\quad\,\,\, \delta D = - \Pi_0, \,\,\,\,\, \delta {\cal P}_D = 0. 
\eea
The partition function is then
\be
\label{5}
Z(\tau) = \int {\cal D}A_\mu \,\, {\cal D}\Pi_\nu \,\, {\cal
  D}(\mbox{ghosts}) \,\, e^{-i S_M(\tau)}, 
\ee
where the Maxwell effective action is 
\be
\label{6}
S_M(\tau) = \int d^4x \,\, \left( \Pi^\mu \dot{A}_\mu + \dot{C} {\cal
  P}_C   + \dot{\cal P}_D D  - H_M - \{ Q, \Psi \}   \right),
\ee
and $\Psi$ is the gauge fixing function. 

As discussed before the Maxwell theory without a theta term and in 
Coulomb gauge has an $SO(2)$ duality symmetry acting on the potentials
\cite{Deser-Teitelboim,Girotti-Gomes-Rivelles}. To go to the Coulomb
gauge we choose $\Psi= \frac{1}{\epsilon}  \partial_i A^i D 
+ A_0 {\cal P}_C$, perform the field
transformation $\Pi_0 \rightarrow \epsilon \Pi_0, \,\, D
\rightarrow \epsilon D$, which has the Jacobian equal to
one,  and take the limit $\epsilon \rightarrow 0$. After integration
over ${\cal P}_C, {\cal P}_D$, $A_0$ and $\Pi_0$ the
partition function Eq.(\ref{5}) reduces to 
\be
\label{7}
Z(\tau) = \int {\cal D}A_i \,\, {\cal D}\Pi_i \,\, {\cal D} C \,\,
{\cal D} D \,\, \delta(\partial_i A^i) \,\, \delta(\partial_i  
\Pi^i) \,\, \exp[ -i \int d^4x \,\, ( \Pi^i \dot{A}_i - H_M +
D \partial^2 C)],
\ee
where $\partial^2 = \partial^i \partial_i$. For the pure Maxwell
theory without a theta term the infinitesimal $SO(2)$ duality
transformations which leave the partition function invariant 
are\footnote{These transformations differ from those of
Eqs.(\ref{1}) by $2\pi$ factors. That is due to different
normalizations for the action.}  
\bea
\label{8}
\delta \Pi_i &=& \frac{\alpha}{2\pi} B_i, \qquad \delta A_i = - 2\pi\alpha
\epsilon_{ijk} \frac{\partial^j}{\partial^2} \Pi^k, \nn
\delta C &=& \delta D = 0.
\eea
On shell they reduce to the usual $SO(2)$ rotation between the
electric and magnetic fields. They also commute with the BRST
transformations. They are local in time but non-local in
space. An important assumption which was implicitly
taken is that the coupling constant $g$ is invariant under duality. In
fact, in \cite{Deser-Teitelboim,Girotti-Gomes-Rivelles} $g^2$ was taken
to be equal to 2.

It is easy to verify that with a theta term the
partition function looses its invariance under duality. In fact, 
the Hamiltonian is no longer invariant. Besides that there are other
annoying points. The first one is that when $\theta = 0$ duality holds
only in Coulomb gauge. A second point is that the coupling constant
changes under duality \cite{Witten} and if we wish to implement the
symmetry on the potentials it should act also on the coupling
constants. However this was not taken into account in
\cite{Deser-Teitelboim,Girotti-Gomes-Rivelles}. The third point
regards the symmetry group. There is, in fact, an $SL(2,\Z)$ symmetry
and not just an $SO(2)$ symmetry 
when a theta term is present \cite{Witten}.  This raises the
question whether it would be possible to implement an $SL(2,\Z)$ symmetry
on the potentials as it was done for $SO(2)$ rotations. 

\section{$SL(2,\Z)$ Duality in Maxwell Theory}
\label{III}

In order to consider the $SL(2,\Z)$ duality it proves to be convenient
to split the vector fields $A_i$ and $\Pi_i$ into their transversal
$A^T_i, \Pi^T_i$ and longitudinal parts $A^L_i, \Pi^L_i$. We will also
consider finite $SL(2,\Z)$ transformations. We have found that the
$SL(2,\Z)$ transformations are given by
\bea
\label{3.1}
&A^T_i& = a \tilde{A}^T_i + 2\pi c \epsilon_{ijk}
\frac{\partial^j}{\partial^2} \tilde{\Pi}^{Tk}, \qquad  \Pi^T_i = d
\tilde{\Pi}^T_i + \frac{b}{2\pi} \tilde{B}_i, \nn
&A^L_i& = |a-c \tilde{\tau}| \tilde{A}^L_i, \qquad\qquad\qquad\,\,  \Pi^L_i =
\frac{1}{|a-c \tilde{\tau}|} \tilde{\Pi}^L_i, \nn
&A_0& = |a-c \tilde{\tau}| \tilde{A}^0, \qquad\qquad\qquad\,\,\,  \Pi_0 =
\frac{1}{|a-c   \tilde{\tau}|} \tilde{\Pi}_0, \nn
&C&= |a-c \tilde{\tau}| \tilde{C}, \qquad\qquad\qquad\,\,\,\,\, {\cal P}_C =
\frac{1}{|a-c \tilde{\tau}|} \tilde{{\cal P}}_C, \nn
&D& = \frac{1}{|a-c \tilde{\tau}|} \tilde{D},
\qquad\qquad\qquad\,\,\,\, {\cal P}_D = |a-c \tilde{\tau}| \tilde{\cal P}_D, 
\eea
\be
\label{3.2}
\tilde{\tau} = \frac{a \tau + b}{c \tau + d},
\ee
where $a,b,c$ and $d$ are integers satisfying $ad-bc=1$.\footnote{If we
  consider just the classical theory the condition that $a,b,c$
  and $d$ are integers can be relaxed and the duality group is then
  $SL(2,\R)$.}  Notice that the transversal (or physical) part of the
vectors are transformed among themselves while the gauge dependent (or
non-physical) parts, which include: $A_0, A_i^L,\Pi_0$ and $\Pi^L_i$
and the ghosts, transform into themselves.  
These transformations are local in time but non-local in space. 

The Hamiltonian density Eq.(\ref{2.2}) is not invariant under
Eqs.(\ref{3.1},\ref{3.2}). It transforms 
as
\bea
\label{3.3}
H_M =  \tilde{H}_M &-& \frac{2\pi i}{\tilde{\tau} - \overline{\tilde{\tau}}}
\frac{2(a - |a-c\tilde{\tau}|) - c(\tilde{\tau} +
  \overline{\tilde{\tau}})}{|a-c\tilde{\tau}|} \tilde{\Pi}^{Ti}
\tilde{\Pi}^L_i \nn
&-& \frac{i}{\tilde{\tau} - \overline{\tilde{\tau}}}
\frac{(a - |a-c\tilde{\tau}|)(\tilde{\tau} + \overline{\tilde{\tau}}) - 2c
  \tilde{\tau}\overline{\tilde{\tau}}}{|a-c\tilde{\tau}|}
\tilde{\Pi}^{Li} \tilde{B}_i,
\eea
and upon integration the extra terms give rise to surface
contributions. The kinetic terms in the effective action Eq.(\ref{6})
are also invariant up to surface terms. Hence, the Hamiltonian is
modular invariant up to surface terms. 

The gauge fixing term in Eq.(\ref{6}) requires some
care. It is easy to show that the BRST charge Eq.(\ref{3}) is
invariant under duality. Hence, it follows that $\Psi$ must be modular
invariant in order that the gauge fixing term in Eq.(\ref{6})
remains modular invariant. Consider first the most general expression
for $\Psi$ which implements linear gauge choices. We write $\Psi$ as 
\be
\label{3.4}
\Psi = \int d^3x \left( \chi D +  A_0 {\cal P}_C \right),
\ee
with $\chi$ depending only on the gauge dependent pieces. If we
require that $\Psi$ be modular invariant under 
duality then $\chi$ must transform as $\chi = |a - c \tilde{\tau}|
\tilde{\chi}$. The most general expression for $\chi$ linear in the
gauge dependent pieces and with the correct transformation property is
\be
\label{3.5}
\chi = \alpha \partial_i A^{Li} + \frac{\beta}{\tau - \overline{\tau}}
\Pi_0 + \gamma A_0 + \frac{\delta}{\tau - \overline{\tau}} \partial^i
\Pi_i^L,
\ee
with $\alpha,\beta,\gamma$ and $\delta$ arbitrary numbers. Therefore
any linear gauge is modular invariant. This includes, among others,
the Coulomb gauge for which $\chi = \frac{1}{\epsilon} 
\partial^i A_i^L$ and covariant gauges for which $\chi = \frac{\xi}{2}
\frac{\Pi_0}{\tau-\overline{\tau}} 
+ \partial^i A_i^L$. It is easy to generalize the above argument for 
nonlinear gauge choices. We have just to multiply the  fields in
$\chi$ by appropriate powers of $\tau - \overline{\tau}$ so that the
product  transforms with a power of $|a-c \tilde{\tau}|$. Hence any
gauge choice can be made modular invariant. 

We then conclude that the effective action is modular invariant for
any gauge  
choice. The Jacobian of the transformations Eqs.(\ref{3.1}) can be
computed and it is found to be equal to one. Therefore, they can be 
regarded as a canonical transformation. Hence the path integral
measure is also invariant. As a consequence, the partition function is
modular invariant. It should be stressed that
the partition function in the Lagrangian formalism is not modular
invariant under duality, rather it transforms as a
modular form \cite{Witten}. However, the phase space partition
function is modular invariant \cite{Lozano,Olive-Alvarez}. 

Finally we must show that Eqs.(\ref{3.1}) reduce to the
familiar duality transformations of the classical equations of
motion. They are given by \cite{Gibbons-Rasheed2} 
\bea
\label{3.6}
G_{\mu\nu} &=& a \tilde{G}_{\mu\nu} + \frac{b}{2\pi} {}^*
\tilde{F}_{\mu\nu}, \nn 
F_{\mu\nu} &=& c {}^* \tilde{G}_{\mu\nu} -\frac{d}{2\pi}
\tilde{F}_{\mu\nu},  
\eea
where $G_{\mu\nu} = -2 \frac{\partial L}{\partial F^{\mu\nu}}$. Also,
  $g^2$ and $\theta$ are identified with the vacuum expectation values
  of the dilaton $\phi$ and axion $a$, respectively, as 
\be
\label{3.7}
\frac{1}{g^2} = \frac{<e^{-\phi}>}{2}, \qquad \frac{\theta}{4\pi^2} =
-<a>. 
\ee
For the Maxwell theory with a theta term we have that 
\be
\label{3.8}
G_{\mu\nu} = -\frac{1}{g^2} F_{\mu\nu} - \frac{\theta}{8\pi^2}
{}^*F_{\mu\nu},
\ee
and we can check that Eqs.(\ref{3.6}) are indeed valid when
Eqs.(\ref{3.1}) are used on-shell, that is, when Gauss law 
holds. Also, the complex combination of the axion and dilaton $a - i
e^{-\phi}$ transforms as $\tau$, as it should. Therefore, the duality
transformations Eqs.(\ref{3.1}) are the off-shell version of
Eqs.(\ref{3.6}).  

At the classical level we can reduce Eqs.(\ref{3.1}) to $SO(2)$
transformations by choosing $a=d=\cos\alpha$ and
$b=-c=\sin\alpha$. For infinitesimal transformations we find that 
\bea
\label{3.9}
\delta A_i^T &=& - 2\pi \alpha \epsilon_{ijk}
\frac{\partial^j}{\partial^2} \Pi^k, \qquad \delta A_i^L = \half \alpha
(\tau + \overline{\tau} ) A_i^L, \nn
\delta \Pi_i^T &=& \frac{\alpha}{2\pi} B_i, \qquad \qquad \qquad 
\delta \Pi_i^L =
\half \alpha (\tau + \overline{\tau} ) \Pi_i^L, \nn
\delta A_0 &=& \half \alpha (\tau + \overline{\tau} ) A_0, \qquad
\,\,\,\,
\delta \Pi_0 = \half \alpha (\tau + \overline{\tau} ) \Pi_0,
\eea
while the real and imaginary parts of $\tau$, respectively $\tau_R$
and $\tau_I$, transform as 
\be
\label{3.10}
\delta \tau_R = - \alpha ( 1 + \tau_R^2 - \tau_I^2 ), \qquad \delta \tau_I =
- 2 \alpha \tau_R \tau_I.
\ee
The transformations for the gauge dependent pieces are proportional to
$\theta$ while the transversal parts have the usual non-local
transformations. When $\theta=0$ the gauge dependent pieces are
invariant and Eq.(\ref{3.10}) fixes the imaginary part of $\tau$ as
$\frac{4\pi}{g^2}=1$. Then the transversal parts have the usual
transformations Eqs.(\ref{8}).

\section{Born-Infeld Theory}
\label{IV}

It is well known that the Born-Infeld theory has an $SO(2)$ symmetry in
its classical equations of motion which can be extended to $SL(2,\R)$
if an axion and a dilaton are added \cite{Gibbons-Rasheed2}. If we
consider just the axion and dilaton vacuum expectation values we
get a Born-Infeld theory with a theta term.  With the identifications
made in Eqs.(\ref{3.7}) its action is 
\be
\label{4.1}
S = \int d^4x \,\, \left( 1 -\frac{\theta}{16 \pi^2}
F^{\mu\nu}{}^*F_{\mu\nu} - \sqrt{1 + \frac{1}{g^2} F^{\mu\nu}
F_{\mu\nu} - \frac{1}{4 g^4} ( F^{\mu\nu} {}^*F_{\mu\nu} )^2 }
\right). 
\ee
In the weak field limit it reduces to Maxwell theory with a theta term
Eq.(\ref{2.1}). There is also a dimensionful constant in the action
which was set equal to one. In order to handle the square root in the
action we introduce an auxiliary field $V$
\be
\label{4.2}
S = \int d^4x \,\, \left[ 1 -\frac{\theta}{16 \pi^2} F^{\mu\nu}
{}^*F_{\mu\nu} - \frac{V}{2} \left( 1 + \frac{1}{g^2} F^{\mu\nu}
F_{\mu\nu} - \frac{1}{4 g^4} ( F^{\mu\nu} {}^*F_{\mu\nu} )^2 \right) -
\frac{1}{2V} \right].  
\ee

The Hamiltonian formulation is straightforward and follows closely
that of \cite{Parra}. Since we have introduced an auxiliary field
$V$ there are two primary constraints $\Pi_0=0$ and
$p=\frac{\partial L}{\partial \dot{V}}=0$. From the first constraint
we get as secondary constraint the Gauss law. From the second
constraint we get an algebraic equation for $V$ which can be solved
so that $V$ is eliminated. We then find the Hamiltonian density 
\be
\label{4.3}
H_{BI} = \sqrt{1 + 2 H_M - (\Pi^i  B_i)^2 + B^i B_i \Pi^j \Pi_j } - 1.
\ee
Clearly the Hamiltonian is not modular invariant under
Eqs.(\ref{3.1},\ref{3.2}). The Maxwell Hamiltonian 
is not invariant and since the duality transformations are linear
the extra terms  in Eq.(\ref{3.3}) can not be
canceled against those coming from $(\Pi B)^2 - B^2 \Pi^2$ term in the
square root in Eq.(\ref{4.3}). Either non-linear terms must be
introduced in Eqs.(\ref{3.1}) or something else must be modified. 

It must be noted that both the Maxwell and the Born-Infeld
Hamiltonian densities can be  rewritten in terms of a complex vector field 
\be
\label{4.4}
P_i = \Pi_i + \frac{\tau}{2\pi} B_i.
\ee
We find that 
\be
\label{4.5}
H_M = - \frac{2\pi i}{\tau - \overline{\tau}} P^i \overline{P}_i,
\ee
and
\be
\label{4.6}
H_{BI} = \sqrt{1 - \frac{4\pi i}{\tau - \overline{\tau}} P^i
\overline{P}_i - \frac{4 \pi^2}{(\tau - \overline{\tau})^2} (P
\times \overline{P})^2 } -1,
\ee
where the overline denotes complex conjugation. The vector $P_i$
transforms under duality as 
\be
\label{4.7}
P_i = \frac{1}{a - c \tilde{\tau}} \left(\tilde{\Pi}^T_i +
\frac{a-c\tilde{\tau}}{|a-c\tilde{\tau}|} \tilde{\Pi}^L_i +
\frac{\tilde{\tau}}{2\pi} \tilde{B}_i \right),
\ee
while 
\be
\frac{1}{\tau - \overline{\tau}} = \frac{ |a - c
  \tilde{\tau} |^2 }{\tilde{\tau} - \overline{\tilde{\tau}}}.
\ee
This explains why the Maxwell Hamiltonian is not invariant. The
longitudinal and transversal parts of $\tilde{\Pi}_i$ do not combine
themselves back into $\tilde{\Pi}_i$ so that $P_i$ is not a modular
form. If instead of $|a-c\tilde{\tau}|$ in the denominator of the
$\tilde{\Pi}^L_i$ term in Eq.(\ref{4.7}) we had just $a-c\tilde{\tau}$
we could recover $\tilde{P}_i$. But taking out the modulus in the
transformations Eqs.(\ref{3.1}) is not consistent because all fields
are real. On the other side if we could change only the transformation
for $\Pi^L_i$ that would do the job. It is then necessary that
$\Pi^L_i$ possess an imaginary part. For consistency $A_0, \Pi_0,
A_i^L$ and the ghosts must have an imaginary part as well.

So we start with the non-physical sector $A_0, \Pi_0, A^L_i, \Pi^L_i$
and the ghosts all described by complex fields. Since the number of ghosts
has also doubled the number of physical degrees of freedom is still
the same. The vectors $A_i$ and $\Pi_i$ are now complex with their
transversal part taken to be real while their longitudinal parts are
taken to be complex. The effective action is now
\be
\label{4.8}
S_{BI} = \int d^4x \,\, \left( \half \overline{\Pi}^\mu \dot{A}_\mu + \half
\Pi^\mu \dot{\overline{A}}_\mu + \half
\dot{C} \overline{\cal P}_C + \half \dot{\overline{C}} {\cal P}_C +
\half \dot{\cal P}_D \overline{D} + \half \dot{\overline{\cal P}}_D
D - H_{BI} - \{Q,\Psi\} \right),
\ee
The Hamiltonian density has the same form as in Eq.(\ref{4.6}) with
$P_i$ defined by Eq.(\ref{4.4}) but with complex fields instead of
real fields. The integrand in the square root in Eq.(\ref{4.6}) is real. 

The BRST charge is now
\be
\label{4.9} 
Q = \half \int d^3x \,\, \left( \partial_i \Pi^i \overline{C} +
\partial_i \overline{\Pi}^i C + \overline{\cal P}_D  \Pi_0 +
{\cal P}_D \overline{\Pi}_0 \right),
\ee
so that $Q$ is real. The BRST transformations are modified in a
straightforward way. The gauge fixing fermion reads now
\be
\label{4.10}
\Psi = \half \int d^3x \,\, \left( \chi \overline{D} + \overline{\chi}
D + A_0 \overline{\cal P}_C + \overline{A}_0 {\cal P}_C \right), 
\ee
and is also real. 

Now we have to show that this theory is equivalent to the original
Born-Infeld theory. In order to do that we will perform a partial
gauge fixing so that all imaginary parts are gauged away. Let us
denote the real and imaginary parts of any complex field $\varphi$ as
$\varphi_R$ and $\varphi_I$, respectively. Let us choose the imaginary part
of $\chi$ as $\chi_I = \frac{1}{\epsilon} \partial_i A^{Li}_I$ and
assume that $\chi_R$ does not depend on $\Pi_{0I}$ and $A_{0I}$. Let
us perform the transformation $\Pi_{0I} \rightarrow \epsilon \Pi_{0I},
D_I \rightarrow \epsilon D_I$, whose Jacobian is equal to one. When
the limit $\epsilon \rightarrow 0$ is taken the
effective action Eq.(\ref{4.8}) reduces to 
\bea
\label{4.11}
&S_{BI}& = \int d^4x \,\, ( \Pi_{0R} \dot{A}_{0R} + \Pi^{Ti} \dot{A}_i^T
+ \Pi^{Li}_R \dot{A}^L_{iR} + \Pi^{Li}_I \dot{A}^L_{iI} + \dot{C}_R
{\cal P}_{CR} + \dot{C}_I {\cal P}_{CI} + \dot{\cal P}_{DR} D_R
- {\cal P}_{DR} {\cal P}_{CR} \nn &-& {\cal P}_{DI} {\cal P}_{CI} - 
\delta \chi_R D_R - C_I \partial^2 D_I - \Pi_{0R} \chi_R -
\Pi_{0I} \partial_i A^{Li}_I + A_{0R} \partial_i \Pi^{Li}_R + A_{0I}
\partial_i \Pi^{Li}_I - H_{BI} ),
\eea
where $\delta\chi_R$ is the BRST transformation of $\chi_R$. Now let
us perform the integral over the imaginary part of all fields. The
integration over $\Pi_{0I}$ produces a delta functional
$\delta(\partial_i A^{Li}_I)$ and since the longitudinal part of
$A_I^i$ has
just one component this means that $A^{Li}_I=0$. Then the integration
over $A^{Li}_I$ can be performed as well. The same is true for the
integration over $A_{0I}$. It gives $\Pi^{Li}_I=0$ and the integration
over $\Pi^{Li}_I$ can also be performed. At this stage $H_{BI}$
depends only on the transversal components $A^T_i, \Pi^T_i$ and the
real part of the longitudinal components $A^L_{iR}, \Pi^L_{iR}$ and
reduces to Eq.(\ref{4.3}). We now integrate over the ghosts. The
integration over ${\cal P}_{CI}$ produces $\delta( \dot{C}_I - {\cal
  P}_{DI} )$ and the integration over ${\cal P}_{DI}$ sets ${\cal
  P}_{DI} = \dot{C}_I$. The remaining integrations over $C_I$ and
$D_I$ gives a $\det \partial^2$ which can be absorbed into the path
integral normalization. Then, only the real part of the ghosts remain
in the effective action and they are the ghost contribution that we
would get if we
had started with all fields real. So we have shown that there is a
gauge choice which eliminates completely the imaginary part of all
fields and that the resulting theory is the original Born-Infeld
theory. 

Then the $SL(2,\Z)$ duality transformations are now 
\bea
\label{4.12}
&A^T_i& = a \tilde{A}^T_i + 2\pi c \epsilon_{ijk}
\frac{\partial^j}{\partial^2} \tilde{\Pi}^{Tk}, \qquad  \Pi^T_i = d
\tilde{\Pi}^T_i + \frac{b}{2\pi} \tilde{B}_i, \nn
&A^L_i& = (a-c \overline{\tilde{\tau}}) \tilde{A}^L_i,
\qquad\qquad\qquad\,  \Pi^L_i = 
\frac{1}{a-c \tilde{\tau}} \tilde{\Pi}^L_i, \nn
&A_0& = (a-c \overline{\tilde{\tau}}) \tilde{A}^0, \qquad\qquad\qquad\,\,
\Pi_0 = \frac{1}{a-c   \tilde{\tau}} \tilde{\Pi}_0, \nn
&C&= (a-c \overline{\tilde{\tau}}) \tilde{C}, \qquad\qquad\qquad\,\,\,\,
{\cal P}_C =
\frac{1}{a-c \tilde{\tau}} \tilde{{\cal P}}_C, \nn
&D& = \frac{1}{a-c \tilde{\tau}} \tilde{D},
\qquad\qquad\qquad\quad\, {\cal P}_D = (a-c \overline{\tilde{\tau}})
\tilde{\cal P}_D.
\eea
The unphysical sector is composed of modular forms. The vector $P_i$
is also a modular form. It transforms as 
\be
\label{4.13}
P_i = \frac{1}{a-c\tilde{\tau}} \tilde{P}_i,
\ee
so that the Maxwell Hamiltonian is modular invariant with no surface
terms being generated. The Born-Infeld Hamiltonian is also modular
invariant. It is easy to show that 
the kinetic terms in the effective action Eq.(\ref{4.8}) are also
invariant up to surface terms. The BRST charge Eq.(\ref{4.9})
is also invariant. By an argument similar to that presented in Section
\ref{III} we conclude that the gauge fixing term in Eq.(\ref{4.8}) is
also modular invariant so that the effective action Eq.(\ref{4.8}) is
modular invariant. Finally we can show that the duality transformations
Eqs.(\ref{4.12}) have a unity Jacobian so that the partition function is
modular invariant. 

The duality transformations Eqs.(\ref{4.12}) reduce to the usual
duality transformations of the classical equations of motion
Eqs.(\ref{3.6}). Now the expression for $G_{\mu\nu}$ is much more
complicated because it involves a square root. However it is
straightforward to show that Eqs.(\ref{3.6}) hold when use is made of
Gauss law. 

\section{Conclusion and Discussion}

We have shown how it is possible to generalize the $SL(2,\R)$ symmetry
of the equations of motion, for Maxwell and Born-Infeld theories, to an
off-shell duality.  For the Maxwell theory we found that the
Hamiltonian $H_M$ is modular invariant up to a surface term. 
In the Born-Infeld case it was necessary to
consider the longitudinal part of the fields as complex fields. Then
the Born-Infeld Hamiltonian $H_{BI}$ is strictly modular 
invariant with no boundary terms being generated by the
transformation. Of course, we could consider Maxwell theory with the
longitudinal part of the fields being complex as well. In this case
the Hamiltonian would be modular invariant without any boundary term. However
there is no clear interpretation for the complex
longitudinal fields introduced in these theories. 

Another important question is whether we can extend the
symmetry to the case where the axion and the dilaton are
propagating fields since it is known that the equations of motion have an
$SL(2,\R)$ symmetry \cite{Gibbons-Rasheed2}. In this case $\tau$ is no
longer constant but a field whose vacuum expectation value is given
by Eq.(\ref{3.7}). The action is then $S = S_0 + S_{BI} $ where 
\be
\label{5.1}
S_0 = -2 \int d^4x 
\frac{\partial^\mu \tau \partial_\mu \overline{\tau}}
{|\tau -  \overline{\tau} |^2}.  
\ee
It is easy to show that $S_0$ is indeed duality invariant. The
Hamiltonian is not modified since no integration by parts was done
and it remains invariant under duality. However, 
the kinetic terms in Eq.(\ref{4.8}) are no longer invariant because
now  $\tau$ is time (and space) dependent. In fact, only the
non-physical sector looses the invariance while the physical one
remains invariant.  

Since self-dual theories, as those studied here, are endowed with
special properties it would 
be interesting to find the supersymmetric extension of the off-shell
duality transformations. There is an intimate connection between
self-duality and spontaneous symmetry breaking of supersymmetry
\cite{Kuzenko} and 
knowing the duality transformations may help to elucidate this
relationship. 

It would be also interesting to study the noncommutative case. Since
$SL(2,\Z)$ is a non-perturbative symmetry of type IIB string theory we
expect that it should be relevant in the noncommutative case as well
\cite{Lu}. However, it seems that the noncommutative gauge theory
obtained from the D3-brane with a B-field along the brane is no longer
self-dual \cite{Ganor}.

It is also known that the equations of motion of p-forms have a
duality symmetry \cite{Lozano}. It would be interesting to find the
extension of our transformations to that case. Another interesting
question is whether our non-local transformations can be made local
along the lines of \cite{Girotti-Gomes-Rivelles}. 

\section{Acknowledgments}
This work is partially supported by FAPESP and PRONEX
No. 66.2002/1998-9. The work of V. O. Rivelles is partially supported
by CNPq. C. R. Le\~ao was supported by a grant from FAPESP.


\begin{references}

\bibitem{Tseytlin} A. A. Tseytlin, ``Self-Duality of Born-Infeld Action
  and Dirichlet 3-brane of type IIB superstring theory´',
  Nucl.Phys. {\bf B469} (1996) 51,  
  hep-th/9602064; M. B. Green and M. Gutperle, ``Comments on
  Three-Branes'', Phys.Lett. {\bf B377} (1996) 28, hep-th/9602077.

\bibitem{Gibbons-Rasheed} G. W. Gibbons and D. A Rasheed,
  ``Electric-Magnetic Duality Rotations in Non-Linear Electrodynamics'',
  Nucl. Phys. {\bf B454} (1995) 185. 

\bibitem{Gaillard-Zumino} M. K. Gaillard and B. Zumino, ``Duality
  Rotations for Interacting Fields'', Nucl. Phys. {\bf B193} (1981)
  221; ``Self-Duality in Nonlinear Electromagnetism'',
  hep-th/9705226. 

\bibitem{Gibbons-Rasheed2} G. W. Gibbons and D. A Rasheed, ``$SL(2,R)$
  Invariance of Non-Linear Electrodynamics Coupled to an Axion and a
  Dilaton'', Phys.Lett. {\bf B365} (1996) 46, hep-th/9509141. 

\bibitem{Deser-Teitelboim} S. Deser and C. Teitelboim, ``Duality
  Transformations of Abelian and Non-Abelian Gauge Fields'', Phys
  Rev. {\bf D13} (1976) 1592.

\bibitem{Bengtsson} I. Bengtsson, ``Manifest Duality in Born-Infeld
  Theory'', Int. J. Mod. Phys. {\bf A12} (1997) 4869, hep-th/9612174. 

\bibitem{Igarashi} Y. Igarashi, K. Itoh and K. Kamimura,
  ``Electric-Magnetic Duality Rotations and Invariance of Actions'',
  Nucl.Phys. {\bf B536} (1998) 454, 
  hep-th/9806160;  ``Self-Duality in Super $D3$-brane Action'',
  Nucl.Phys. {\bf B536} (1998) 469, 
  hep-th/9806161. 

\bibitem{Chern-Simons} D. Birmingham, M. Rakowski and  G. Thompson,
  ``Renormalization of Topological Field Theory'', Nucl. Phys. {\bf
  B329} (1990) 83.

\bibitem{Damgaard} P. H. Damgaard and V. O. Rivelles,
  ``Symmetries of Chern-Simons Theory in Landau Gauge'', Phys. Lett. 
{\bf 245 B} (1990) 48. 

\bibitem{Schwarz-Sen} D. Zwanziger, ``Local Lagrangian Quantum Field
  Theory of Electric and Magnetic Charges'', Phys. Rev. {\bf D3}
  (1971) 880; J. H. Schwarz and A. Sen, ``Duality Symmetric
  Actions'', Nucl. Phys. {\bf B411} (1994) 35, hep-th/9304154.

\bibitem{McClain} B. McClain, Y. S. Wu and F. Yu, ``Covariant
  Quantization of Chiral Bosons and $OSP(1,1|2)$ Symmetry'',
  Nucl. Phys. {\bf B343} (1990) 689; I. Bengtsson and A. Kleppe, ``On
  Chiral p-Forms'', 
  Int. J. Mod. Phys. {\bf A12} (1997) 3397,  hep-th/9609102;
  N. Berkovits, ``Local Actions with Electric and Magnetic Sources'',
  Phys. Lett. {\bf B395} (1997) 28, hep-th/9610134.

\bibitem{Girotti-Gomes-Rivelles} H. O. Girotti, M. Gomes,
  V. O. Rivelles and A. J. Silva, ``Duality Symmetry in the
  Schwarz-Sen Model'', Phys. Rev. {\bf D56} (1997) 6615, hep-th/
  9702065.

\bibitem{PST} P. Pasti, D. Sorokin and M. Tonin, ``Duality Symmetric
  Actions with Manifest Space-Time Symmetries'', Phys.Rev. {\bf D52}
  (1995) 4277, hep-th/9506109.

\bibitem{Parra} D. Berman, ``$SL(2,\Z)$ Duality of Born-Infeld Theory
  from Nonlinear Selfdual Electrodynamics in Six-Dimensions'',
  Phys. Lett. {\bf B409} (1997) 153, hep-th/9706208; A. Khoudeir and
  Y. Parra, ``On Duality in the Born-Infeld Theory'', Phys. Rev. {\bf
  D58} (1998) 025010, hep-th/9708011. 

\bibitem{Rivelles} V. O. Rivelles, ``Comment on ``A New Symmetry for
  QED'' and  ``Relativistically Covariant Symmetry in QED'''',
  Phys. Rev. Lett. {\bf 75} (1995) 4150, hep-th/9509028; ``Several
  Guises of the BRST Symmetry'', Phys. Rev. {\bf D53} (1996) 3247,
  hep-th/9510136.

\bibitem{Witten} E. Witten, ``On S-Duality in Abelian Gauge Theory'',
  hep-th/9505186; E. Verlinde, ``Global Aspects of Electric-Magnetic
  Duality'',  Nucl.Phys. {\bf B455} (1995) 211, hep-th/9506011. 

\bibitem{Lozano} Y.Lozano, ``S-Duality in Gauge Theories as a
  Canonical Transformation'', Phys.Lett. {\bf B364} (1995) 19,
  hep-th/9508021; ``Duality and Canonical Transformations'',
  Mod. Phys. Lett. {\bf A11} (1996) 2893, hep-th/9610024. 

\bibitem{Olive-Alvarez} D. I. Olive and M. Alvarez, ``Spin and Abelian
  Electromagnetic Duality on Four-Manifolds'', hep-th/0003155.

\bibitem{Kuzenko} S. M. Kuzenko and S. Theisen, ``Nonlinear
  Self-Duality and Supersymmetry'', hep-th/0007231.

\bibitem{Lu} J. X. Lu, S. Roy and H. Singh, ``SL(2, Z) Duality and
4-Dimensional Noncommutative Theories'', hep-th/0007168.

\bibitem{Ganor} O. Ganor, G. Rajesh and S. Sethi, ``Duality and
  Non-Commutative Gauge Theory'', Phys.Rev. {\bf D62} (2000) 125008,
  hep-th/0005046; S-J. Rey and R. von Unge, ``S-Duality, Noncritical
  Open String and Noncommutative Gauge Theory'', hep-th/0007089. 

\end{references}
\end{document}